\documentclass[a4paper,notitlepage]{revtex4}
\usepackage[pdftex]{graphicx}
\usepackage{times}
\usepackage{amsfonts}
\usepackage[active]{srcltx}
\usepackage{color}

\begin{document}

\title{On the analogy between a single atom and an optical resonator}

\author{S. Heugel$^1$}
\email{simon.heugel@mpl.mpg.de}
\author{A. S. Villar$^1$}
\author{M. Sondermann$^{1,2}$}
\author{U. Peschel$^{1,2}$}
\author{G. Leuchs$^{1,2}$}
\email{gerd.leuchs@mpl.mpg.de}

\affiliation{ $^{1}$Max Planck Institute for the Science of Light,
              G\"unther-Scharowsky-Str. 1/Bau 24,
              91058 Erlangen, Germany\\
              $^{2}$Institute of Optics, Information and Photonics,
              University of Erlangen-Nuremberg,
              Staudtstr. 7/B2,
              91058 Erlangen, Germany}

\begin{abstract}
A single atom in free space can have a strong influence on a light beam and a 
single photon can have a strong effect on a single atom in free
space. 
Regarding this interaction, two conceptually 
different questions can be asked: can a single atom fully absorb a single photon 
and can a single atom fully reflect a light beam. 
The conditions for achieving the full effect in either case are
different. 
Here we discuss related questions in the context of an optical resonator. 
When shaping a laser pulse properly it will be fully absorbed by an
optical resonator, i.e., no light will be reflected and all the pulse
energy will accumulate inside the resonator before it starts leaking out. 
We show in detail that in this case the temporal pulse shape has to
match the time-reversed pulse obtained by the cavity's free decay. 
On the other hand a resonator, made of highly reflecting mirrors which
normally reflect a large portion of any incident light, may fully
transmit the light, as long as the light is narrow band and resonant
with the cavity.
The analogy is the single atom - normally letting most of the light 
pass - which under special conditions may fully reflect the incident light
beam. 
Using this analogy we are able to study the effects of practical
experimental limitations in the atom-photon coupling, such as finite
pulses, bandwidths, and solid angle coverage, and to use the
optical resonator as a test bed for the implementation of the quantum
experiment.
\end{abstract}

\maketitle 
\section{Introduction}

The strong interaction in free space between a single quantum of light and an individual 
two-level quantum system is a fundamental physical process. 
The first observation of the partial extinction of a light beam by a single
ion dates back to 1987 \cite{wineland1987}.
Nevertheless, the topic is of broad ongoing interest and there is a
wealth of recent papers related to it, covering experiments
\cite{vamivakas2007,wrigge2008,tey2008-np} as well as theoretical and
conceptual works
\cite{vanenk2001,vanenk2004,sondermann2007,pinotsi2008,tey2008-a,zumofen2008p,stobinska2009}. 

The terms \emph{extinction} applies to two qualitatively
different situations:
A strongly focused, coherent, low intensity continuous wave (cw) light
beam on resonance with the two-level quantum system -- which we
exemplarily assume to be an atom here -- can be scattered by the atom.
The probability to find it in the excited state remains small during the whole interaction procedure.
This case is treated in Refs.~\cite{vamivakas2007,wrigge2008,tey2008-np,tey2008-a,zumofen2008p}.
Carrying this scenario to extremes, it is predicted that a cw beam focused from half the solid angle onto the atom is completely reflected back into the half space from where it originated \cite{zumofen2008p}.
In other words, the cw beam is completely extincted.  
On the contrary, the field incident onto the atom may constitute of a light pulse containing only a single photon. 
Now, the extreme case is that there is a distinct moment in time where the probability to find the atom in the excited state is unity and the incident photon is fully absorbed (and re-emitted after wards).
This approach is followed in Refs.~\cite{sondermann2007,pinotsi2008,stobinska2009}.

It is common to both approaches that the spatial properties (angular intensity dependence, state of polarization) of the focused light must match the atomic dipole transition in order to achieve the desired maximum effect~\cite{quabis2000,vanenk2001,vanenk2004,lindlein2007,pinotsi2008,zumofen2008p}.
However, this is not enough for the aim of unit absorption probability.
Here, the incident photon must possess the time reversed properties of a photon spontaneously emitted by the atom~\cite{quabis2000,lindlein2007,stobinska2009}.
This requirement naturally includes the conditions for the spatial profile of the photon mentioned above but also sets restrictions on the temporal and spectral properties: Since the spontaneous decay is exponential \cite{weisskopf1930}, the temporal envelope of the incident photon has to be an increasing exponential (cf. Ref. \cite{stobinska2009}) up to a point in time where it drops off rapidly.
Furthermore, since the decay occurs into the full solid angle the atom must be illuminated from all directions.

The differences and similarities between these two situations can be better envisaged by tracing an analogy to a well-known classical system, a light beam resonant to an optical cavity. 
The excitation probability of the atom is then associated to the energy present in the light field internal to the optical resonator, and the conditions for the complete absorption, reflection, or transmission of the incident classical field in this situation can be used to better understand the absorption or reflection of photons by a two-level quantum system.
The analogy is of course only approximate because contrary to the atom the quantization of the field plays no role in the treatment of the classical resonator.

The atom--cavity analogy will be established as follows:
We will assume that the light incident onto the resonator is perfectly
mode matched to it.
This means that we consider all requirements on the spatial intensity
distribution and polarization pattern to be fulfilled.
Then, the discussion can be focused onto spectral and temporal issues and
the amount of the solid angle covered by the incident light.
With these settings the pulsed excitation from full solid angle is
compared to a resonator that is irradiated through all its mirrors.
For the sake of simplicity, we will assume a linear resonator comprising only of two mirrors, one perfectly and one partially reflecting. 
The effect of half solid angle illumination of an atom finds its counterpart in a resonator consisting of two partially reflecting mirrors with equal reflectivity illuminated only from one side.
Following the reasoning of Ref.~\cite{zumofen2008p} one can then introduce two half--spaces: One where the light directly reflected by the cavity front mirror interferes with the light leaking out of the resonator backwards through the front mirror and one which contains only light leaking out of the resonator, i.e., the light transmitted by the resonator.
When introducing the similarity between the resonator and the atom the only subtlety is the need to associate reflection by the resonator
with transmission by the atom.

In this manner, we are able to study the response of a two-level quantum system upon single photon excitation, i.e., its excitation and  the temporal properties of the scattered radiation, by analyzing the analogous response of a simple classical system. 
We show that the only way to completely store all the incident pulse energy inside the resonator is to shape its temporal profile as the phase-conjugate of the  wave emitted by the freely decaying system.
We also show that for the symmetric resonator excited through only one of the two input mirrors the maximum energy storage is 50\% of the energy of the incident time-inverted pulse. 
This allows us to conjecture that the two-level atom irradiated from half of the total 
solid angle will likewise reach an excitation of at most 50\%.
In the continuous-wave case and limiting the effective field excitation to a single photon the totally reflecting atom is indeed equivalent to an optical resonator with equal mirrors  for which all resonant light is transmitted.

\section{Comparison of the experimental schemes}

\begin{figure}[h]
 \includegraphics[width=8cm]{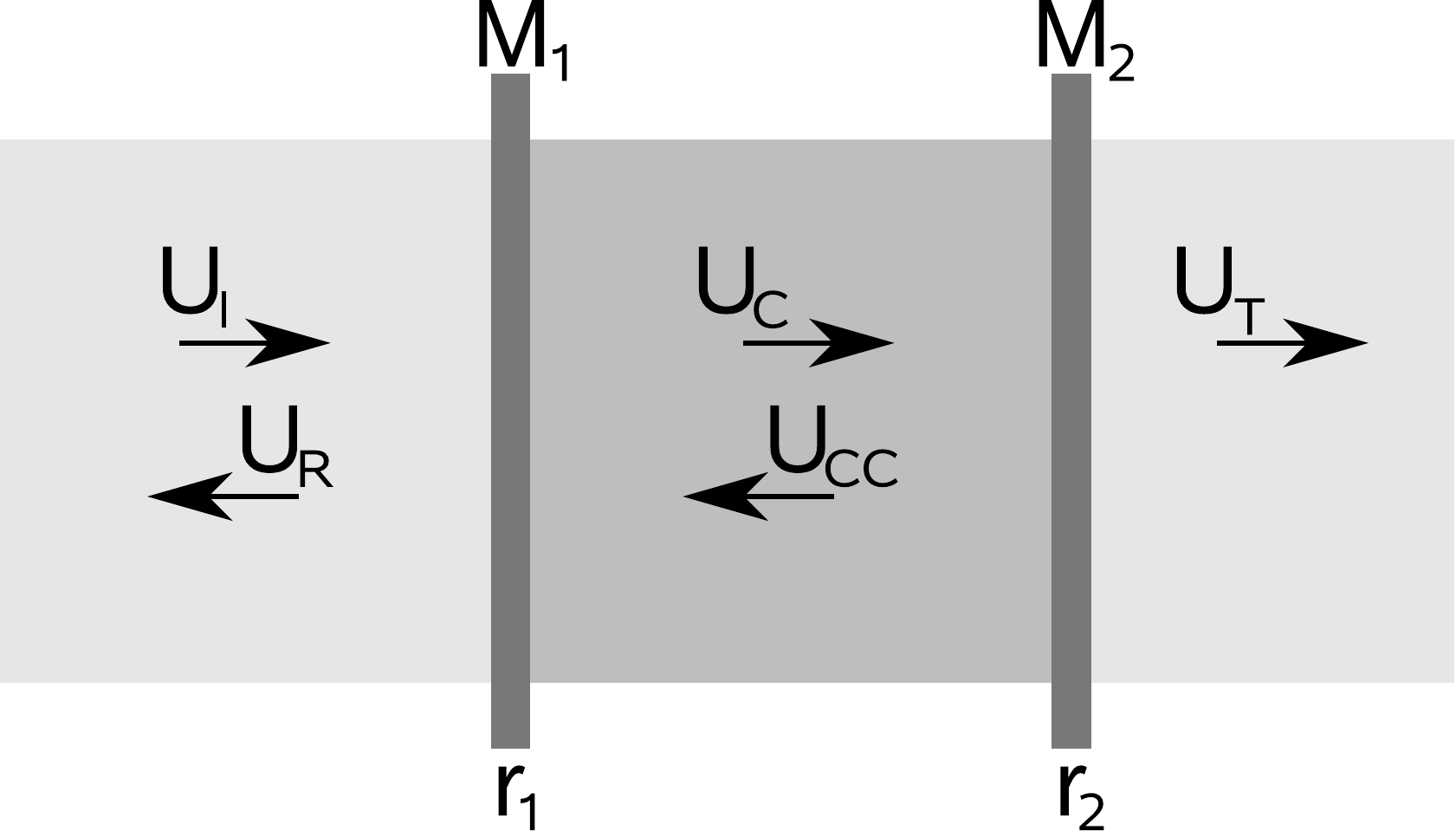}
 \caption{Scheme of a simplified Fabry-P\'erot resonator made of two parallel semi-transparent mirrors. In the calculations, only mirror $M_1$ is illuminated.}
 \label{fig:Resonator}
\end{figure}

The analogy between atoms and optical resonators can be found in the limit of high finesse for the latter.
The optical resonator is designed as a Fabry-P\'erot cavity, created from two plane parallel mirrors of infinite size.
The normal mode of such a simplified resonator are plane waves, such that spatial and time domain parts of these normal modes separate according to $\vec E(\vec r,t)=U(t)\vec E_0(\vec r)$. The resonator dynamics are fully described by the time domain part $U(t)$ while only one polarization component is considered here.
This time domain part is chosen such, that the intensity is given by $I(t)=\left|U(t)\right|^2$.

The setup of such a cavity is shown in Fig.~\ref{fig:Resonator}.
The phase reference for all the light beams is taken at the position of mirror $M_1$.
The phase-shifts regarding reflection at the mirror surfaces are set in such a way, that the reflection of light at the outside border of each mirror gives rise to a phase shift of $\pi$.

The light beam being incident onto the cavity is $U_I$.
The cavity is only irradiated from one side.
The light being reflected by the cavity is described by the component $U_R$ and the transmitted component is $U_T$.
Inside the cavity the two components $U_C$ and $U_{CC}$ circulate.
The resonator has a fixed length $L$ so that the photon round trip time is defined as $\tau=2L/c$.
The resonator's resonances are given by $\omega_n=n\nu_{fsr}$ with $n\in\mathbb N$ and the free spectral range $\nu_{fsr}=1/\tau$.

The dynamics of the cavity being excited by incident light beams is best described in the frequency domain \cite{cesini1977}.
For constant mirror positions and due to it's linear response, the cavity acts as a linear filter on the spectrum of the incident light pulse.
The incident pulse is defined as $U_I(t)=p(t)e^{i\omega_0t}$, the spectrum of which reads as
\begin{equation}
\label{eq:IncidentFieldSpectrum}
 U_I(\omega)=p(\omega-\omega_0).
\end{equation}
The reflected and transmitted fields are obtained from the product of this incident spectrum with the corresponding 
filter functions, given for an arbitrary optical resonator by
\begin{equation}
  \label{eq:CavityReflection}
  C_R(\omega)= \frac{-r_1+r_2e^{-i\omega/\nu_{fsr}}}{1-r_1r_2e^{-i\omega/\nu_{fsr}}} 
\end{equation}
for the reflection and by
\begin{equation}
  \label{eq:CavityTransmission}
    C_T(\omega)=
  \frac{t_1t_2e^{-i\omega/(2\nu_{fsr})}}{1-r_1r_2e^{-i\omega/\nu_{fsr}}} 
\end{equation} 
for the transmission. The internal field decay rate is 
\begin{equation}
\label{eq:DecayRate}
\Gamma=-\nu_{fsr}\ln(r_1r_2).
\end{equation}

The energy stored in the cavity is finally obtained by the expression
\begin{equation}
  \mathcal E(t)=\mathcal A
  \int\limits_{-\infty}^{t}\!\left[\left|U_I(t')\right|^2-
  \left|U_R(t')\right|^2-\left|U_T(t')\right|^2\right]\mathrm{d}t' 
\end{equation}
with the cavity area $\mathcal A$.
By normalizing this quantity to the total energy of the incident field 
$\mathcal{E_P}=\mathcal A\int\limits_{-\infty}^{\infty}\left|U_I(t')\right|^2\mathrm{d}t'$,
the fraction of energy $\epsilon(t)$ being stored in the cavity at time $t$ can be 
simply calculated as 
$\epsilon(t)=\mathcal{E}(t)/\mathcal{E_P}$. 
It is useful to define the energy absorption efficiency $\epsilon_{max}=\max_{t\in\mathbb R} \epsilon(t)$.

In those cases where the quantities $U_T,U_R$ have been computed, these were obtained by taking the product of the filter functions with the incident light pulse $U_I$ in discrete Fourier space \cite{cesini1977,shakir1983,christodoulides1986,yu2001}.
The free spectral range was set to $\nu_{fsr}=10$~GHz and the reflectivity 
of the coupling mirror $M_1$ to $r_1=0.9999$. We used the discrete Fourier transform
to numerically compute the frequency domain components, where the sampling 
frequency was set to $5~\nu_{fsr}$ and the duration of the time window to more 
than $500~\Gamma^{-1}$.

\subsection{Perfect absorption of a matching light pulse}

The case of perfect absorption of light by an atom in free space corresponds in our picture to the compression of a light pulse into a single-ended resonator ($r_2=1$).
The total absorption of the matching light pulse into the resonator can be understood as a suppression of reflection from the resonator during the time the pulse is irradiated onto mirror $M_1$.
The same picture applies to an atom which is irradiated from the full solid angle with a focused spatial pattern which matches that of the internal transition which is driven (e.g. a dipolar pattern).
There the light which is emitted from the partially excited atom destructively interferes with the light which has passed the atoms position.

\begin{figure}[h]
  \includegraphics[angle=270,totalheight=3.8cm]{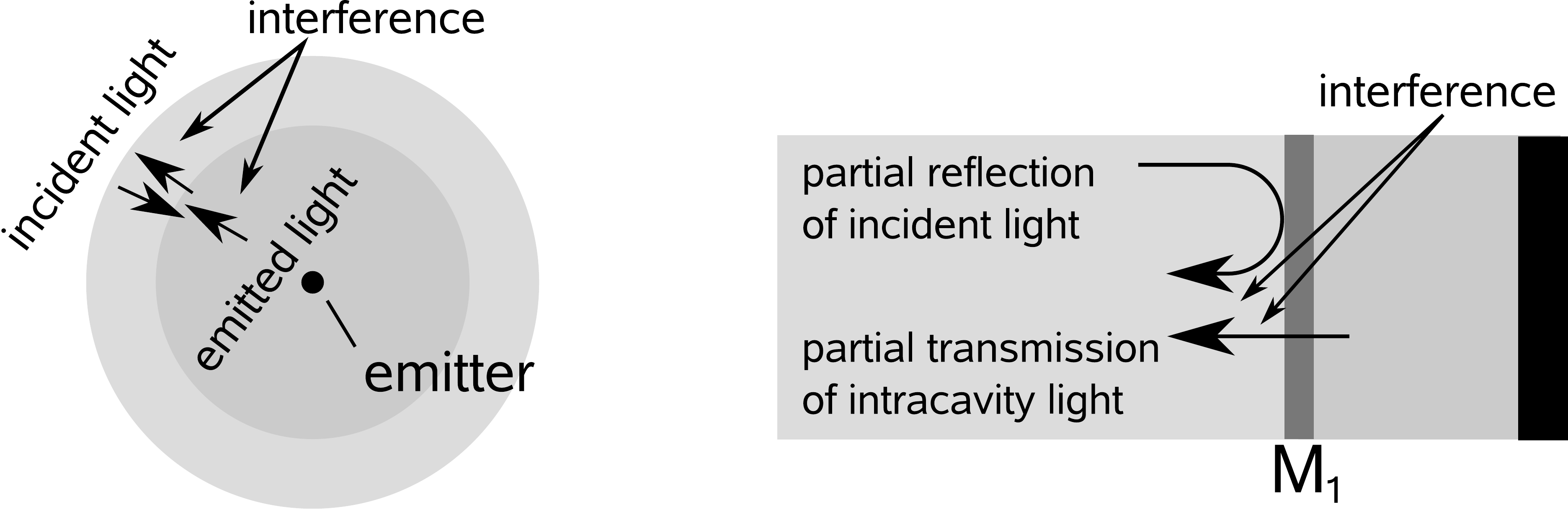}
  \caption{Comparison of the interferences of light beams in the case of complete illumination of the systems.}
  \label{fig:FullSolidAngleComparison}
\end{figure}

First, the equation governing the resonator response onto arbitrary pulse shapes are rewritten to accommodate the special symmetry of such a single-ended resonator.
From the fact that there is no transmission, it can be seen that such a resonator acts as a dispersive mirror \cite{christodoulides1986}.
Only frequency dependent phase shifts are imposed on the spectrum of the incident light pulse.
We note that this property can be used for instance to passively convert phase-squeezed light into 
amplitude-squeezed light, and vice-versa \cite{Galatola199195,villar07}.

The reflection filter function in Eq.~\ref{eq:CavityReflection} can therefore be described fully by its argument $\phi(\omega)\nolinebreak=\nolinebreak\arg \left\lbrace \widetilde{C_R}(\omega) \right\rbrace$.
Due to the high finesse of the cavity, $\frac{\omega-\omega_0}{\nu_{fsr}}\ll1$ holds for all relevant frequencies being involved in here. So $\phi(\omega)$ can be approximated as
\begin{equation}
 \phi(\omega)\approx\arctan \left[ - \frac{2(1-r_1)\frac{\omega-\omega_0}{\nu_{fsr}}}{(1-r_1)^2-\left(\frac{\omega-\omega_0}{\nu_{fsr}}\right)^2} \right]+\pi.
\end{equation}
This reduces the complete cavity filter function to the form of
\begin{equation}
 \widetilde{C_R}(\omega)=\frac{\Gamma - i(\omega-\omega_0)}{\Gamma + i(\omega-\omega_0)}.
\end{equation}

At this point, the application of time reversal regarding the optimal pulse becomes obvious.
Consider a pulse which grows exponentially with the rate $\Gamma$ until it suddenly stops at $t=0$.
Such a pulse
\begin{equation}
\label{eq:ExponentialGrowingPulseTimeDomain}
p(t)=p_0e^{\Gamma t}H(-t)
\end{equation}
with the Heaviside step function $H(t)$ and an amplitude $p_0$ has a spectrum 
\begin{equation}
\label{eq:ExponentialGrowingPulseFreqDomain}
\widetilde{p}(\omega-\omega_0)=\frac{p_0}{\Gamma-i(\omega-\omega_0)}.
\end{equation}
So the spectrum of the reflected light beam is
\begin{equation}
\label{eq:ExponentialGrowingReflectionSpectrum}
\widetilde{U_R}(\omega)=\frac{p_0}{\Gamma+i(\omega-\omega_0)}.
\end{equation}
This reflected light beam is the time inverse of the incident light beam due to 
$\widetilde{U_R}(\omega)=\widetilde{U_I}(\omega)^*$.

Going back to the time domain picture, the total concentration of the incident light pulse into the cavity becomes immediately obvious.
The reflected component reads
\begin{equation}
 \label{eq:ReflectedComponentTimeDomain}
 U_R(t)=p_0e^{(i\omega_0-\Gamma) t}H(t)
\end{equation}
so until $t=0$ no light is leaving the cavity.
This is due to a destructive interference between the light which leaves the cavity during that time, and the light which is directly reflected at mirror $M_1$ (Fig. \ref{fig:FullSolidAngleComparison}).
As $U_R(t)$ vanishes during the time the pulse is incident onto the cavity, all its energy flows into the cavity region and is confined there until $t=0$, hence perfect absorption of the pulse energy by the cavity is accomplished at $t=0$.

It has to be emphasized here that the symmetry of the source--free Maxwell equations under the time reversal transformation $t\mapsto -t$ is the basic reason for this remarkable total absorption of light energy into a generally lossy system.
By irradiating the resonator with the time reversal of that pulse which it would emit while decaying freely, this system is driven to behave in a time-reversed way.

\subsection{Perfect reflection (resp. transmission) of a narrow band faint light beam}

\begin{figure}[h]
  \includegraphics[angle=270,totalheight=3.8cm]{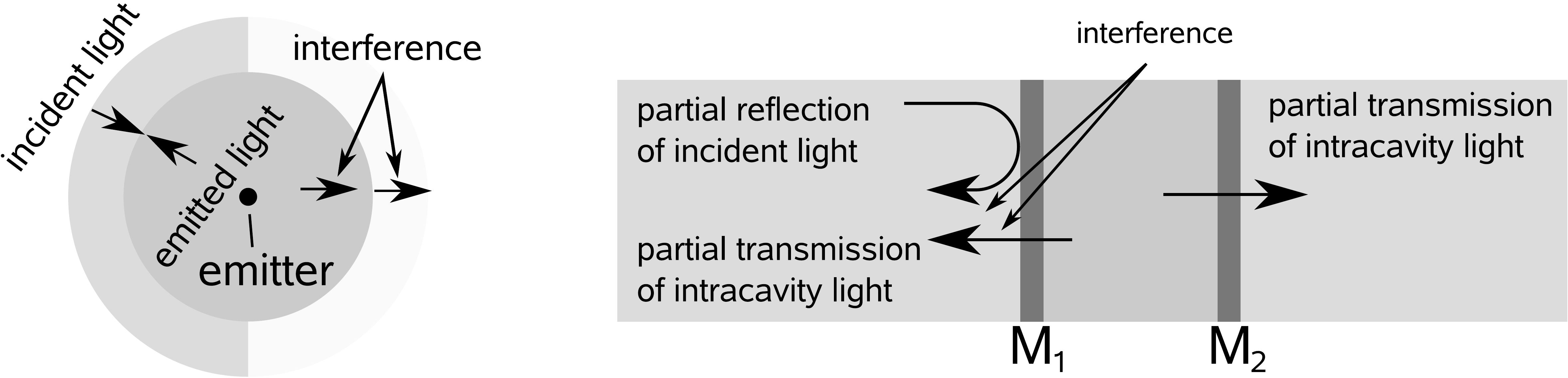}
  \caption{Comparison of the interference effects which lead to full transmission (resonator) or full reflection (atom) in the case of resonant excitation with faint light beams.}
  \label{fig:HalfSolidAngleComparison}
\end{figure}

The action of a symmetric double-ended cavity ($r_1=r_2$) such as in Fig. \ref{fig:Resonator} on narrow band resonant light is well known from text-books.
It can also be seen from Eq. \ref{eq:CavityReflection} that in resonance $C_R(n\nu_{fsr})=0$ for $n\in\mathbb N$.
Therefore all light is transmitted in this special case, independent of the finesse.

This remarkable effect is due to the interference of the light which is partially reflected off mirror $M_1$ and the light which leaks out through this cavity mirror from inside the resonator $\sqrt{1-r_1^2}U_{CC}$.
This interference determines the total reflected light $U_R$.
In the case of resonance, these two components are balanced with opposite signs.
Therefore, all the light energy is transmitted through the resonator.

A direct analog of this perfect transmission of resonant narrow band light through a symmetric resonator is the total reflection of a narrow band faint light beam from a single quantum emitters.
The same picture of partial destructive interference of beams can be drawn, if the atom is irradiated from one half of the full solid angle with a focused spatial pattern which matches that of the internal transition which is driven (e.g. a dipolar pattern).
The portion of incident light which passes the position of the atom interferes with the light emitted by the atom.
If this interference is fully destructive, as is the case for resonant light beams, all the incident light is reflected back from the atom.

The reason for the interchange of transmission and reflection in both cases can be seen from direction in which interference occurs.
This is visualized in Fig. \ref{fig:HalfSolidAngleComparison} where schemes of both experimental situations are shown.

\subsection{Comparison of other experimental schemes regarding the absorption efficiency}

\subsubsection{Absorption of a narrow band faint light beam into a single-ended resonator}

\begin{figure}[h]
  \resizebox{0.45\textwidth}{!}{\includegraphics[angle=270]{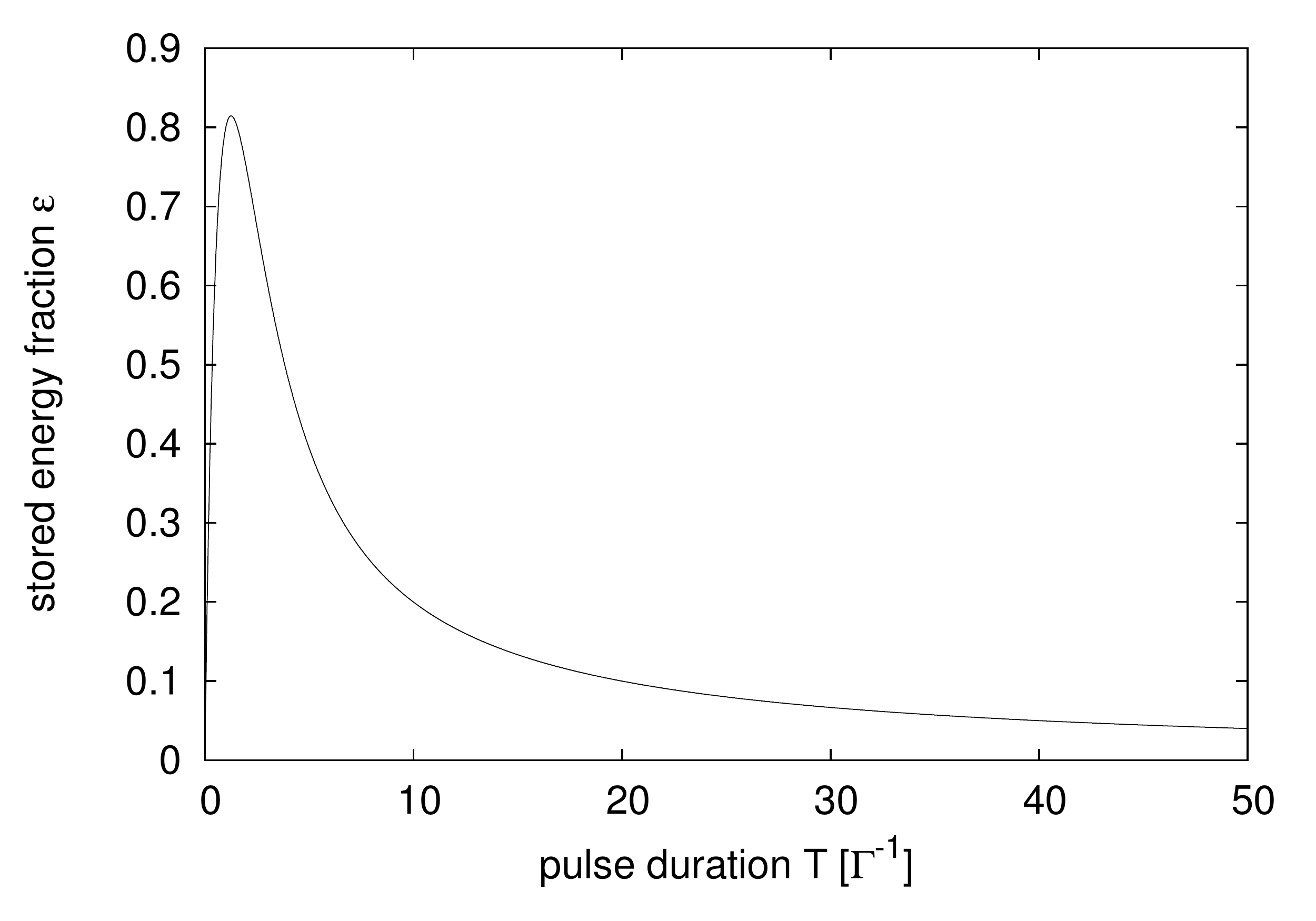}}
  \caption{Absorption efficiency $\epsilon_\mathrm{max}$ as a function of the
  pulse length $T$ for a rectangular pulse shape.}
  \label{fig:RectangularLaserPulse}
\end{figure}

In our simulations we approximate narrow-band light beams by pulses of rectangular shape with a sufficient duration, i.e., the spectral width of these pulses is much smaller than the width of the cavity resonance.
In order to investigate the transition from short pulses to narrow band light beams, we computed the absorption efficiency dependence on the pulse length utilizing the above introduced technique.
The results in Fig. \ref{fig:RectangularLaserPulse} indicate that only for short pulse durations a considerable fraction of energy can be stored inside the resonator.
This is due to the effect that in the beginning transient signals impose the build up of energy storage inside the resonator.
Once the steady state is reached, no more energy will be stored inside the resonator.
If the duration of the rectangular pulse is longer than this initial build-up process takes, the fraction of energy being stored inside the resonator decreases with increasing pulse durations.
For the atomic experiments with illumination from half the solid angle this means, that although perfect reflection occurs for a narrow band faint light beam, no significant excitation of the atom will be observed.

\subsubsection{Partial absorption of a matching light pulse into a symmetric double-ended resonator}

\begin{figure}[h]
  \resizebox{0.45\textwidth}{!}{\includegraphics[angle=270]{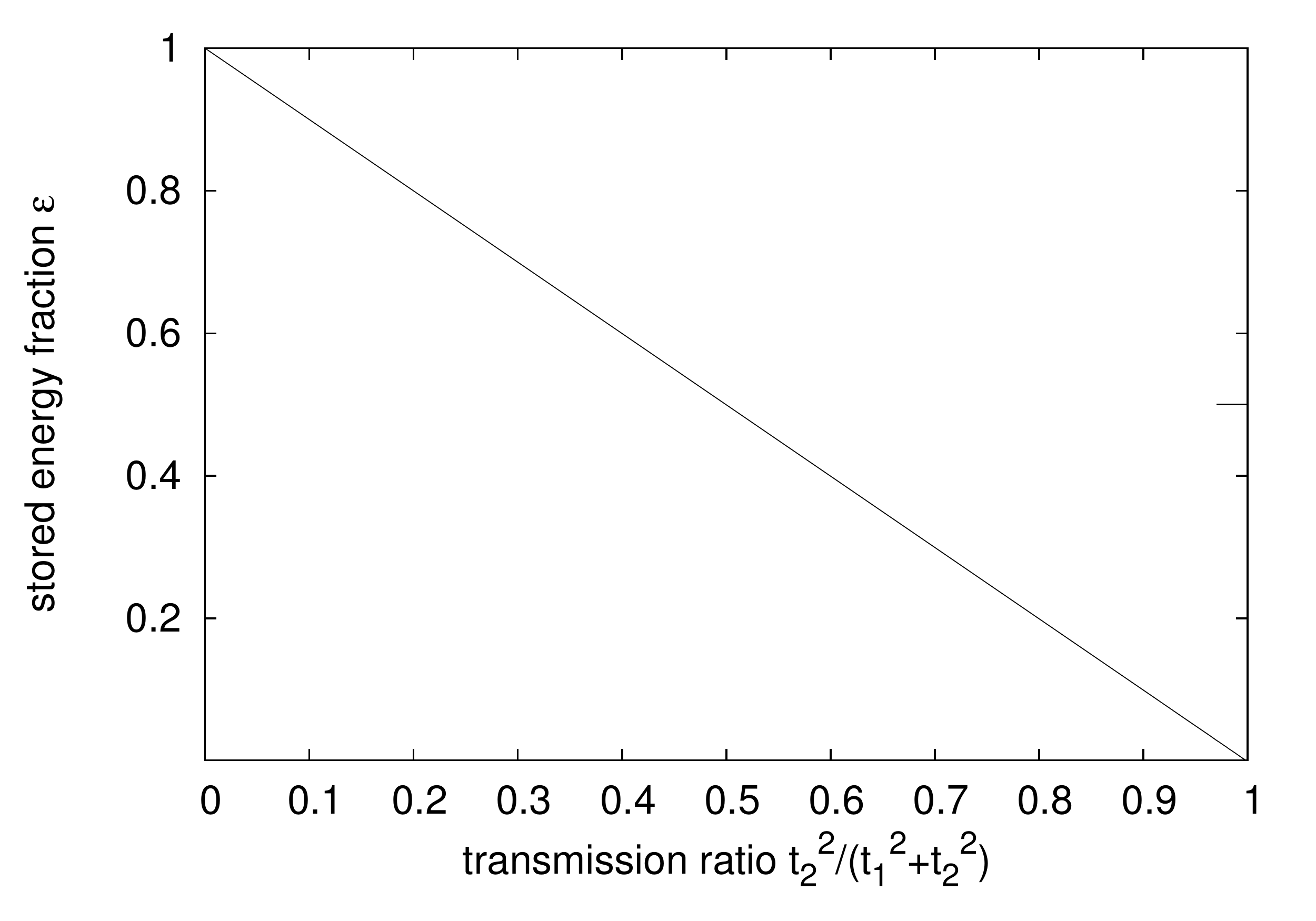}}
  \caption{Absorption efficiency $\epsilon_{max}$ as a function of the relative transmission of energy through mirror $M_2$.}
  \label{fig:AbsorptionSymmResonator}
\end{figure}

The best strategy for finding the optimum pulse shape seems to be to match the decay constant $\Gamma$ of the particular resonator.
While the resonator is irradiated, the reflection of light is canceled.
Therefore the only loss regarding  energy storage is due to light-leakage through mirror $M_2$.
The results in Fig. \ref{fig:AbsorptionSymmResonator} show that for symmetric resonators where $t_2^2/(t_1^2+t_2^2)=0.5$, the maximum of fractional energy storage in the resonator amounts to $\epsilon_{max}=0.5$.
According to the analogy between both systems, this limitation to $\epsilon_{max}=0.5$ also applies to all experiments where atoms are illuminated from only half of the full solid angle.
 
\section{Effect of imperfections on the total absorption of light energy by optical resonators}

\subsection{Deviations from the matching pulse shape}

We now consider how realistic pulses would modify the ideal dynamics discussed in the previous section. 
The influence of various deviations from the optimal pulse shape on the achievable absorption efficiency 
is investigated numerically.
The reflectivity of mirror $M_2$ has thereby been set to $r_2=1$.

\begin{figure}[h]
  \resizebox{0.45\textwidth}{!}{%
    \includegraphics[angle=270,width=8.5cm]{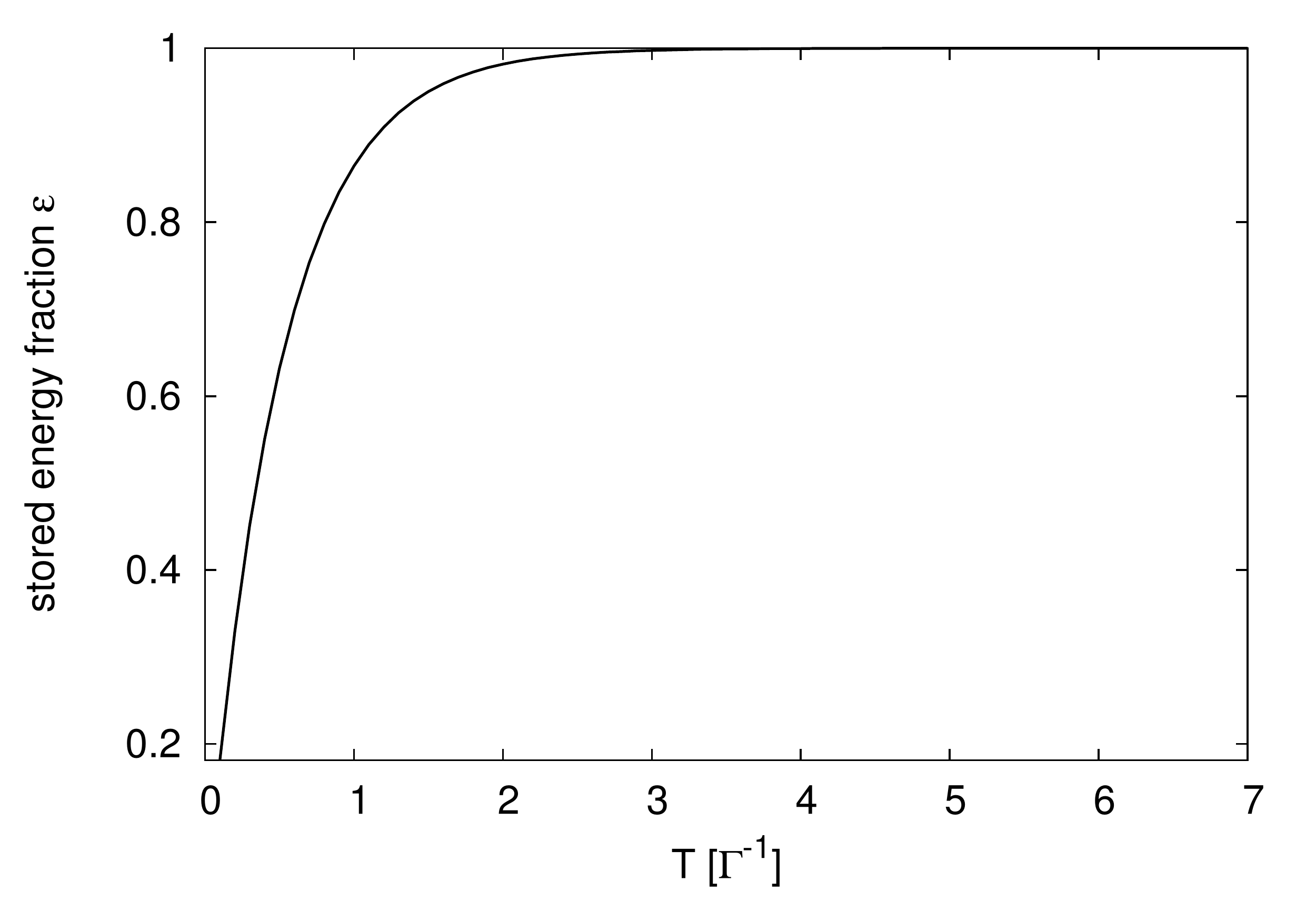}
  }
  \caption{The absorption efficiency $\epsilon$ as a function of the
  pulse length $T$ for a pulse shape $p(t)=p_0e^{\Gamma t}H(-t)H(t+T)$.}
  \label{fig:FinitePulses}
\end{figure}

Realistic pulses will be restricted to a finite length $T$, being written as
\begin{equation}
 \label{eq:FinitLengthPulse}
 p(t)=p_0e^{\Gamma t}H(-t)H(t+T),
\end{equation}
where $H(t)$ is the Heaviside step function. 
Fig.~\ref{fig:FinitePulses} depicts the influence of finite pulse durations on the
amount of energy stored in the cavity. It becomes clear that relatively short 
pulses of duration $T=4\Gamma^{-1}$ are already absorbed with an efficiency as high 
as $\epsilon_\mathrm{max}=0.9997$. The reason why finite $T$ implies $\epsilon_\mathrm{max}<1$ 
can be explained by the partial reflection of the very initial pulse tail, which will always be reflected without
interfering destructively with the internal field, since this yet has to be built up.
But due to the exponentially growing shape of the incident pulses, this effect is 
negligible if the initial step is small enough. 

\begin{figure}[h]
  \resizebox{0.45\textwidth}{!}{%
    \includegraphics[angle=270,width=8.5cm]{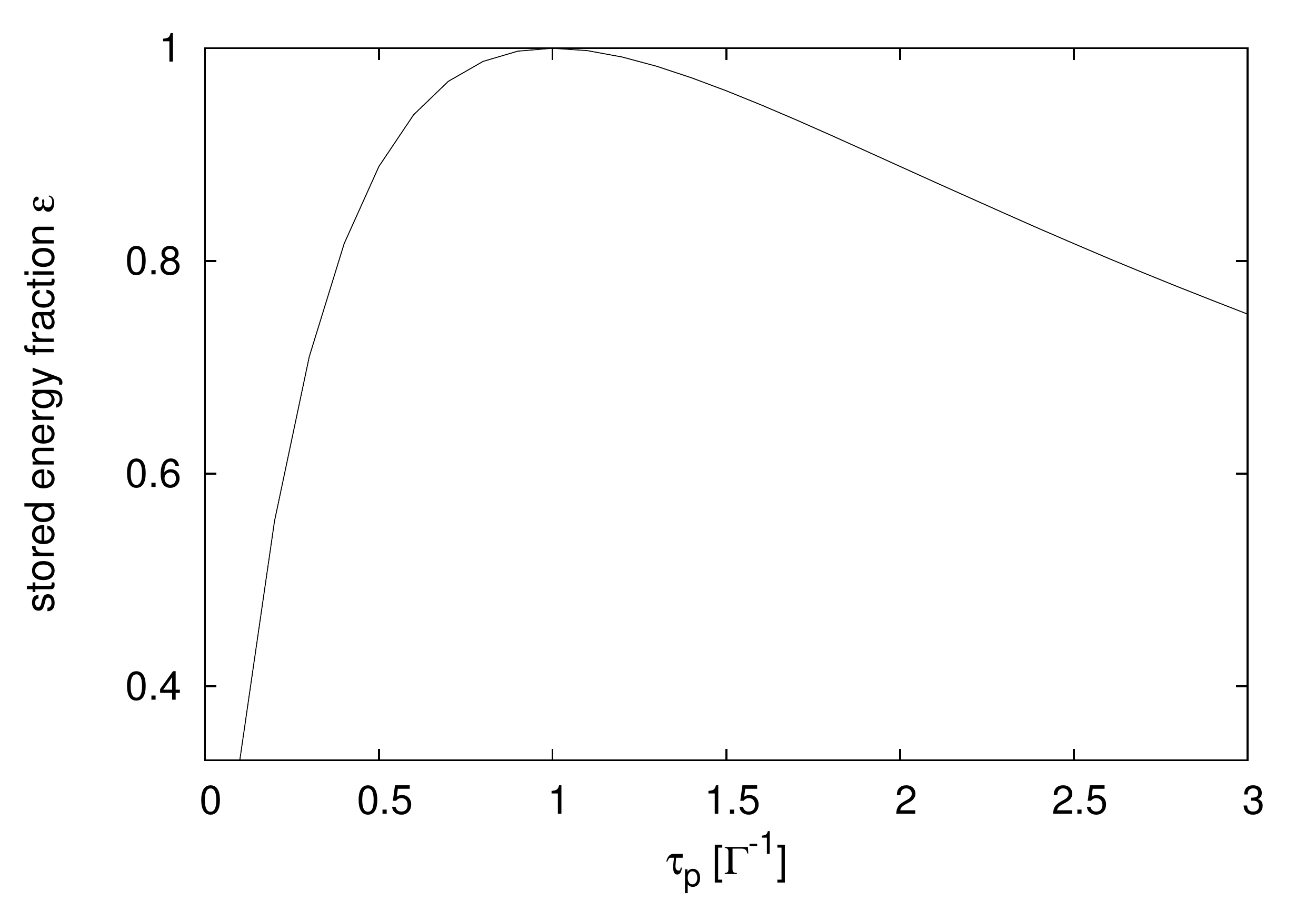}
  }
  \caption{Absorption efficiency $\epsilon_\mathrm{max}$ as a function of the
  exponential time constant $\tau_p$ for an incident pulse shape $p(t)=p_0e^{t/\tau_p}H(-t)$.}
  \label{fig:ExpRate}
\end{figure}

Realistic pulses will also suffer from a second problem, the mismatch between
its time constant decay $\tau_p$ and the cavity lifetime $\Gamma$, breaking once more
the ideal symmetry required. The results are presented in Fig. \ref{fig:ExpRate},
where $\epsilon_\mathrm{max}$ is plotted as a function of the incident pulse 
exponential time constant. 
The flat maximum seen at $\tau_p=\Gamma^{-1}$ clearly shows that small deviations 
of $\tau_p$ do not lead to a sharp drop in excitation efficiency, showing that 
the generation of the incident pulse is robust also against this kind of error.

\subsection{Resonator losses through $r_2<1$}

\begin{figure}[h]
\resizebox{0.45\textwidth}{!}{%
  \includegraphics[angle=270,width=8.5cm]{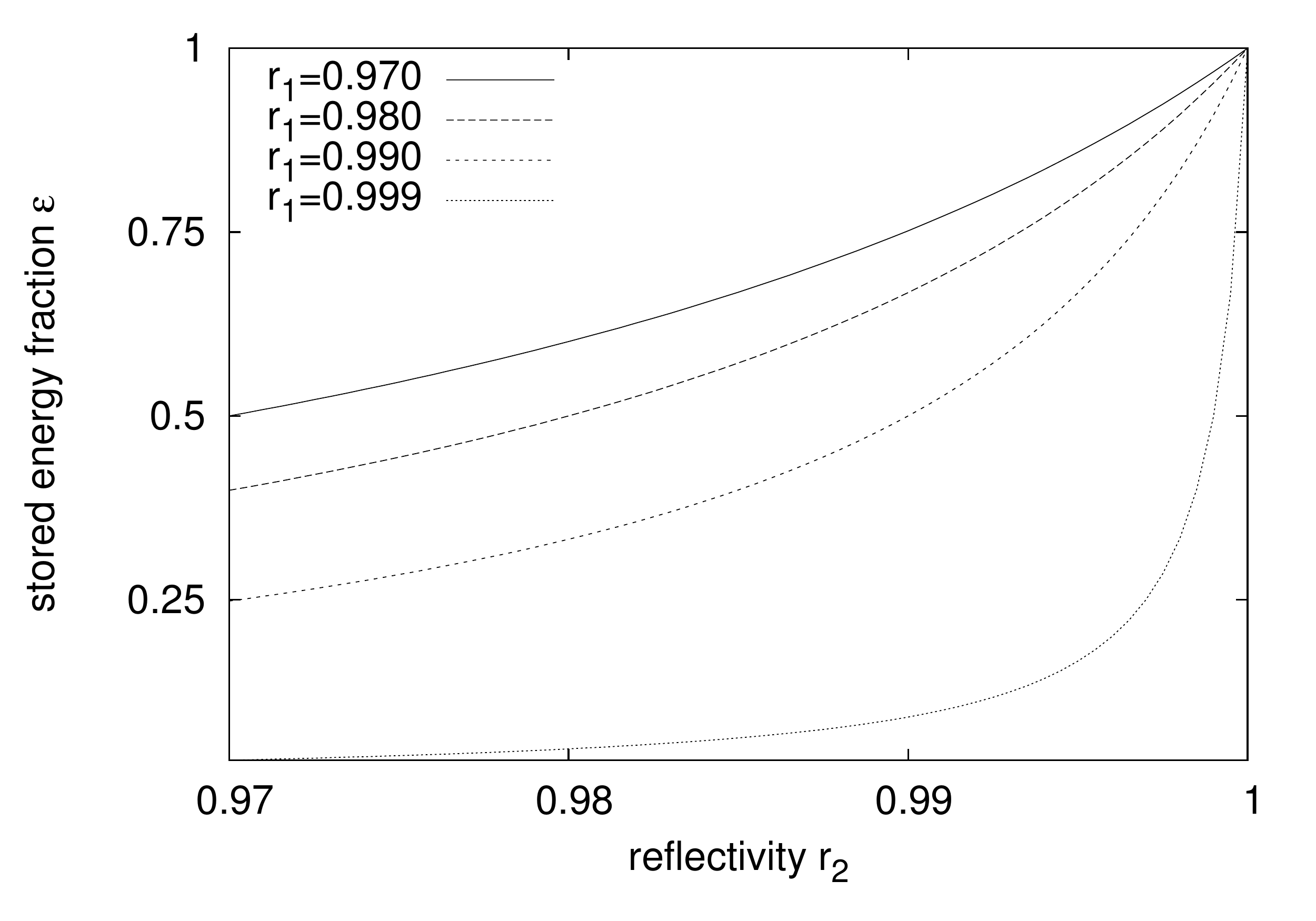}
}
\caption{The effect of losses through mirror $M_2$ onto the absorption efficiency
  $\epsilon$ for $r_1=(0.97,0.98,0.99,0.999)$.} 
\label{fig:ScanReflectivity}
\end{figure}

The simplification $r_2=1$ that has been applied so far can hardly be fulfilled in any experiment.
Therefore, the case of $r_2<1$ is investigated next.
The results in Fig. \ref{fig:ScanReflectivity} show how strong the absorption efficiency depends on the prevention of loss through mirror $M_2$.
There is a non-linear drop off in absorption efficiency if $r_2<1$.
This can be explained as follows:
While the light pulse is incident onto the cavity, the intracavity field builds up and circulates inside the cavity.
Due to the high finesse, it passes mirror $M_2$ many times.
Even small transmissions, compared to those through mirror $M_1$, accumulate and have a considerable effect.

\section{Conclusions}

We are pointing at a remarkable similarity between the interaction of light with an atom and with an optical resonator. 
As discussed the similarity holds in several respects.
This may seem surprising because it links the dynamics of the linear optical resonator at any incident light power level with the dynamics of an atom interacting with up to one photon.
However, if the atom is exposed to higher photon fluxes, the similarity breaks down. 
The reasons are that the atomic energy spectrum is not harmonic and that nonlinear effects become important. 
Because of this similarity one has a useful test bed at hand for investigating experimental strategies for the preparation of the pulses to be employed in the atom experiment and to estimate deficiencies resulting from imperfect light fields.  
We concentrated on two cases, one where the linear resonator is excited through both open ports, and the other where it is excited only through one of the two open ports.
For the atom this corresponds to the cases where the incident focused light field matches a dipole wave in the full solid angle and one where it matches the dipole wave in a limited solid angle cone $<4\pi$.
For modelling the situation with the resonator one just has to choose the corresponding reflectivities $r_1$ and $r_2$.
The interaction of the atom with $n\leq1$ photons can thus be quantitatively simulated using the resonator model.
However, this similarity does not at all cover the efficient coupling of the atom in free space at higher photon numbers where novel possibilities for non linear optics at the few photon level may become feasible \cite{Nemoto2004,stobinska2009,turchette1995}.

\begin{acknowledgements}
The authors gratefully acknowledge fruitful discussions with Magdalena Stobi\'nska.
\end{acknowledgements}

\end{document}